\documentclass[aps,prl,superscriptaddress,twocolumn,showpacs,amsmath,amssymb,letter]{revtex4}

\usepackage{graphicx}
\usepackage{dcolumn}
\usepackage{bm}

\begin{document}

\title{Large Gap Topological Insulator Bi$_2$Te$_3$ with a Single Dirac Cone on the Surface}

\author{Y. L. Chen}
\affiliation {Stanford Institute for Materials and Energy Sciences,
SLAC National Accelerator Laboratory, 2575 Sand Hill   Road, Menlo
Park, California 94025} \affiliation {Geballe Laboratory for
Advanced Materials, Departments of Physics and Applied Physics,
Stanford University, Stanford, California 94305} \affiliation
{Advanced Light Source, Lawrence Berkeley National Laboratory
Berkeley California, 94720, USA }

\author{J. G. Analytis}
\affiliation {Stanford Institute for Materials and Energy Sciences,
SLAC National Accelerator Laboratory, 2575 Sand Hill   Road, Menlo
Park, California 94025} \affiliation {Geballe Laboratory for
Advanced Materials, Departments of Physics and Applied Physics,
Stanford University, Stanford, California 94305}

\author{J. H. Chu}
\affiliation {Stanford Institute for Materials and Energy Sciences,
SLAC National Accelerator Laboratory, 2575 Sand Hill   Road, Menlo
Park, California 94025} \affiliation {Geballe Laboratory for
Advanced Materials, Departments of Physics and Applied Physics,
Stanford University, Stanford, California 94305}

\author{Z. K. Liu}
\affiliation {Stanford Institute for Materials and Energy Sciences,
SLAC National Accelerator Laboratory, 2575 Sand Hill   Road, Menlo
Park, California 94025} \affiliation {Geballe Laboratory for
Advanced Materials, Departments of Physics and Applied Physics,
Stanford University, Stanford, California 94305}

\author{S. K. Mo}
\affiliation {Geballe Laboratory for Advanced Materials, Departments
of Physics and Applied Physics, Stanford University, Stanford,
California 94305} \affiliation {Advanced Light Source, Lawrence
Berkeley National Laboratory Berkeley California, 94720, USA }

\author{X. L. Qi}
\affiliation {Stanford Institute for Materials and Energy Sciences,
SLAC National Accelerator Laboratory, 2575 Sand Hill   Road, Menlo
Park, California 94025} \affiliation {Geballe Laboratory for
Advanced Materials, Departments of Physics and Applied Physics,
Stanford University, Stanford, California 94305}

\author{H. J. Zhang}
\affiliation {Beijing National Laboratory for Condensed Matter
Physics, and Institute of Physics, Chinese Academy of Sciences,
Beijing 100190, China }

\author{D. H. Lu}
\affiliation {Stanford Institute for Materials and Energy Sciences,
SLAC National Accelerator Laboratory, 2575 Sand Hill   Road, Menlo
Park, California 94025}

\author{X. Dai}
\affiliation {Beijing National Laboratory for Condensed Matter
Physics, and Institute of Physics, Chinese Academy of Sciences,
Beijing 100190, China }

\author{Z. Fang}
\affiliation {Beijing National Laboratory for Condensed Matter
Physics, and Institute of Physics, Chinese Academy of Sciences,
Beijing 100190, China }

\author{S. C. Zhang}
\affiliation {Stanford Institute for Materials and Energy Sciences,
SLAC National Accelerator Laboratory, 2575 Sand Hill   Road, Menlo
Park, California 94025} \affiliation {Geballe Laboratory for
Advanced Materials, Departments of Physics and Applied Physics,
Stanford University, Stanford, California 94305}

\author{I. R. Fisher}
\affiliation {Stanford Institute for Materials and Energy Sciences,
SLAC National Accelerator Laboratory, 2575 Sand Hill   Road, Menlo
Park, California 94025} \affiliation {Geballe Laboratory for
Advanced Materials, Departments of Physics and Applied Physics,
Stanford University, Stanford, California 94305}

\author{Z. Hussain}
\affiliation {Advanced Light Source, Lawrence Berkeley National
Laboratory Berkeley California, 94720, USA }

\author{Z. X. Shen}
\affiliation {Stanford Institute for Materials and Energy Sciences,
SLAC National Accelerator Laboratory, 2575 Sand Hill   Road, Menlo
Park, California 94025} \affiliation {Geballe Laboratory for
Advanced Materials, Departments of Physics and Applied Physics,
Stanford University, Stanford, California 94305}

\date{\today}

\begin{abstract}
We investigate the surface state of Bi$_2$Te$_3$ using angle
resolved photoemission spectroscopy (ARPES) and transport
measurements. By scanning over the entire Brillouin zone (BZ), we
demonstrate that the surface state consists of a single
non-degenerate Dirac cone centered at the $\Gamma$ point.
Furthermore, with appropriate hole (Sn) doping to counteract
intrinsic n-type doping from vacancy and anti-site defects, the
Fermi level can be tuned to intersect only the surface states,
indicating a full energy gap for the bulk states, consistent with a
carrier sign change near this doping in transport properties. Our
experimental results establish for the first time that Bi$_2$Te$_3$
is a three dimensional topological insulator with a single Dirac
cone on the surface, as predicted by a recent theory.

\end{abstract}


\maketitle

Soon after the theoretical prediction\cite{Bernevig1}, a new state
of quantum matter - the two-dimensional (2D) topological insulator
displaying the quantum spin Hall(QSH) effect - was experimentally
observed in the HgTe quantum wells\cite{Konig}. The QSH
state\cite{Kane1,Bernevig2} has an insulating gap in the bulk and
gapless states at the edge where opposite spin states
counter-propagate. The two opposite spin states form a single
massless Dirac fermion at the edge, and the crossing of their
dispersion branches at a time reversal invariant point is protected
by the Kramers theorem. This robust protection is a consequence of
the $Z2$ topological quantum number of the bulk quantum
states\cite{Kane2}. The dissipationless edge state transport of the
QSH state may enable low power spintronics devices.

A few years ago, 2D massless Dirac fermions were experimentally
discovered in graphene with two inequivalent massless Dirac points
for each spin orientation, giving rise to four copies of massless
Dirac fermions in total. This is consistent with the experimentally
observed quantized Hall conductance in units of
$2e^2/h$\cite{Zhang1}, as each Dirac fermion leads to a quantized
Hall conductance in units of $e^2/(2h)$ in an external magnetic
field. It is no accident that graphene has an even number of
massless Dirac fermions, since no time-reversal invariant purely 2D
fermion system can have a single, or an odd number of massless Dirac
fermions. Therefore, one can only observe a single Dirac fermion in
a 2D system if it is the boundary of a three-dimensional (3D)
system, called a 3D topological insulator\cite{Fu1,Moore,Roy,Qi1}. A
3D topological insulator has a bulk insulating gap with gapless
surface states inside the bulk gap. In the simplest case, the
surface state consists of a single Dirac cone, with one quarter the
degrees of freedom of graphene. In this case, the electrodynamics of
the topological insulator is described by an additional topological
term in the Maxwell's equation\cite{Qi1}, leading to striking
quantum phenomena such as an image magnetic monopole induced by an
electric charge\cite{Qi2}.

The 3D material HgTe under strain is predicted to have a single
Dirac cone on the surface\cite{Dai}. However, experiments are
difficult to perform under the strain condition. The $Bi_{1-\delta}
Sb_\delta$ alloy is also predicted to be a 3D topological insulator
in the narrow alloying content regime of
$\delta=0.07\sim0.22$\cite{Fu2,Teo}, and a recent ARPES study
reveals the novel nature of the surface state despite its
complexity, with as many as five branches crossing the Fermi level
($E_F$)\cite{Hsieh}. However, the small bulk gap
($\sim10meV$\cite{Lenoir} or $\sim50meV$\cite{Hsieh}) makes it
vulnerable to intrinsic random substitutional disorder, possibly
leading to bulk impurity bands which can overlap with the surface
states or smearing out of the bulk gap. Considering the lack of
systematic doping dependent measurements, it is unclear whether the
material is indeed insulating in the bulk. Furthermore, the small
gap  system is not well suited for realistic high(room) temperature
applications.

Recently, a new class of stoichiometric materials, $Bi_2Te_3$,
$Bi_2Se_3$ and $Sb_2Te_3$ has been theoretically predicted to be 3D
topological insulators whose surface states consist of a single
Dirac cone at the $\Gamma$ point\cite{Zhang2}. In a recent ARPES
experiment\cite{Xia} on $Be_2Se_3$, a single surface electron pocket
with a Dirac point below the Fermi level at the $\Gamma$ point was
reported. However, a deep bulk electron pocket coexisting with the
topologically non-trivial surface states was also observed in the
same experiment. Therefore, the topological insulating behavior in
this class of materials is yet to be established experimentally.

In this letter, we use ARPES and transport experiments to
investigate both the bulk and surface state electronic properties of
$(Bi_{1-\delta}Sn_\delta)_2Te_3$ crystals (where $\delta$ represents
nominal Sn  concentration, incorporated to compensate n-type doping
from vacancy and anti-site defects. By scanning over the entire
Brillouin zone, we confirm that the surface states consist of a
single, non-degenerate Dirac cone at the $\Gamma$ point. At
appropriate doping ($\delta=0.67\%$), we found that the bulk states
disappear completely at the Fermi level, thus realizing for the
first time the topological insulating behavior in this class of
materials. With a much larger bulk band gap ($165meV$) compared to
the energy scale of room temperature($24meV$), the topological
protection of the surface states in this material could lead to
promising applications in low power spintronics devices at room
temperature.

\begin{figure}
\includegraphics[width=0.5\textwidth]{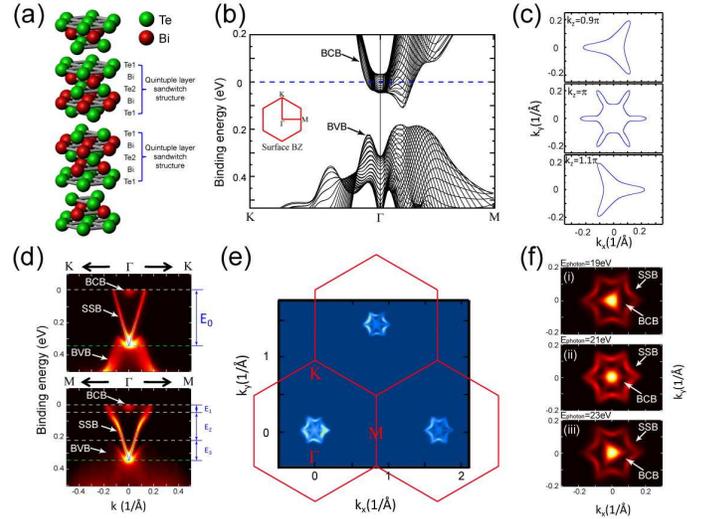}
\caption{\label{fig:epsart} (Color) Crystal and electronic
structures of $Bi_2Te_3$ (a) Tetradymite-type crystal structure of
$Bi_2Te_3$, formed by stacking quintuple-layer groups sandwiched by
three sheets of Te and two sheets of Bi. (b) Calculated bulk
conduction band(BCB) and bulk valance band(BVB) dispersions along
high symmetry directions of the surface BZ (see inset), with the
chemical potential rigidly shifted to 45meV above the BCB bottom at
to match the experimental result.  (c) The $k_z$ dependence of the
calculated bulk FS projection on the surface BZ. (d) ARPES
measurements of band dispersions along $K-\Gamma-K$(top) and
$M-\Gamma-M$ (bottom) directions. The broad bulk band (BCB and BVB)
dispersions are similar to those in panel (b), while the sharp
V-shape dispersion is from the surface state band (SSB). Energy
scales of the band structure are labeled as: $E_0$: Binding energy
of Dirac point (0.34eV), $E_1$: BCB bottom binding energy(0.045eV),
$E_2$:bulk energy gap(0.165eV) and $E_3$: energy separation between
BVB top and Dirac point (0.13eV). (e) Measured wide range FS map
covering three BZs shows that the FSs only exist around $\Gamma$
point, where the red hexagons represent the surface BZ. The uneven
intensity of the FSs at different BZs results from the matrix
element effect. (f) Photon energy dependent FS maps. The shape of
the inner FS changes dramatically with photon energies, indicating a
strong $k_z$ dependence due to its bulk nature as predicted in panel
(c), while the non-varying shape of the outer hexagram FS confirms
its surface state origin.}
\end{figure}

Fig. 1 summarizes the bulk and surface electronic structures and
Fermi-surfaces (FSs) topology of undoped $Bi_2Te_3$. The crystal
structure of $Bi_2Te_3$ (Fig. 1a) is of the tetradymite type, formed
by stacking quintuple layer groups sandwiched by three sheets of Te
and two sheets of Bi within each group\cite{Wyckoff}. $Ab$ $initio$
calculations predict that the undoped $Bi_2Te_3$ is a semiconductor
(Fig. 1b), and that the Fermi-surface(Fig. 1c) from bulk conduction
band(BCB) projected onto the surface Brillouin zone($BZ$) exhibits
triangular or hexagonal snowflake-like electron pocket centered at
the $\Gamma$ point (Fig. 1c) depending on its $k_z$ position in
reciprocal space.

Actual band dispersions measured by the ARPES experiment along two
high symmetry directions are shown in Fig 1d. In addition to the
broad spectra of the bulk electron pocket on top and the ``M" shape
valance band at bottom as predicted in the $ab$ $initio$
calculation, there is an extra sharp ``V" shape dispersion resulted
from the surface state. The linear dispersion in both plots clearly
indicates a massless Dirac Fermion with a velocity of
$4.05\times10^5m/s (2.67 eV\cdot\AA)$ and $3.87\times10^5m/s (2.55
eV\cdot\AA)$ along the $\Gamma-K$ and $\Gamma-M$ directions
respectively, which are about $40\%$ of the value in
graphene\cite{Zhang1} and agree well with our first principle
calculation (by the method described in\cite{Zhang2}) which yields
2.13 and 2.02 $eV\cdot\AA$ along the $\Gamma-K$ and $\Gamma-M$
directions, respectively.

This sharp surface state also forms a FS pocket in addition to the
calculated FS from bulk bands. As shown in Fig. 1e, 1f(ii), in each
BZ there is a hexagram FS enclosing the snowflake like bulk FS. A
broad FS map covering three adjacent BZs (Fig. 1e) confirms that
there is only one such hexagram FS resulting from the ``V" shape
Dirac-type surface state in each BZ. It should be noted that the
spin-orbit coupling (SOC) in this material is rather strong and the
atomic SOC of Bi-6p orbital is $\lambda=1.25eV$\cite{Zhang2}, about
twice of that in Au ($\lambda=0.68eV$)\cite{Vijayakumar}. Given that
our energy and momentum resolution ($\delta E<0.016eV$ and $\delta
k<0.012(1/\AA)$) is better than needed to resolve even the much
smaller Au surface state splitting ($\Delta E=0.11eV, \Delta
k=0.023(1/\AA)$)\cite{LaShell},  The fact that we never observe more
than one set of surface state in all dopings and under all
experimental conditions - including different photon energies,
polarizations and two experimental setups in different synchrotrons
-- rules out the possibility that the Dirac cone is spin
degenerated. This crucial observation clearly demonstrates that
$Bi_2Te_3$ is the ideal candidate as the parent compound for the
simplest kind of 3D topological insulator\cite{Zhang2} - a
simplicity resembling that of the hydrogen atom in atomic physics.
In contrast, graphene has two valleys with spin degeneracy, totaling
four Dirac cones in each BZ, leading to a topological trivial state.
Furthermore, since there is only one surface Fermi pocket in each
surface BZ, surface state will only cross the Fermi level $once$
between ¦£ and M, rather than the complex crossing of  $five$ times
as observed in $Bi_{0.9}Sb_{0.1}$\cite{Kane1, Hsieh}.

The surface nature of the hexagram FS resulting from the sharp ``V"
shape dispersion is further established by a photon energy
dependence study (Fig. 1f). By varying the excitation photon energy,
the shape of the snowflake-like bulk FS changes from a left pointing
triangle(Fig. 1f(i)) to a right pointing triangle(Fig. 1f(iii)) as a
result of the $k_z$ dispersion of the 3D bulk electronic structure
as illustrated in Fig. 1c. In contrast, the shape of the
hexagram-like FS does not change with the incident photon energy,
confirming its two dimensional nature ($i.e.$ no $k_z$ dispersion).
We note that the perfect $Bi_2Te_3$ single crystal is predicted to
be a bulk insulator. The electron carriers observed in our
experiment arise from crystal imperfections, specifically vacancies
and anti-site defects \cite{Saiterthwaite}. Given the substantial
bulk gap (Fig. 1d), one can tune the $E_F$ into the gap by doping
holes to compensate the electron carriers, thus realizing the
topological insulator phase in this material. In order to lower the
Fermi level into the bulk gap, we introduce controlled Sn doping
into $Bi_2Te_3$. As a Sn atom has one less valence electron than a
Bi atom, substituting Bi by Sn effectively dopes holes into the
compound, which decreases the bulk electron density and results in a
downshift of $E_F$\cite{Kulbachinskii}. The effect of Sn doping is
clearly demonstrated in Fig. 2, where the FSs and band dispersions
of samples of four different nominal dopings ($0\%, 0.27\%, 0.67\%$,
and $0.9\%$) are shown from panel (a) to (d), respectively.  The top
row shows the evolution of the FSs. The surface state FS pocket is
observed in all four compounds, whose volume shrinks with increasing
doping, with its shape varying from a hexagram ($0\%, 0.27\%$ and
$0.67\%$ dopings) to an hexagon ($0.9\%$ doping). The evolution of
the bulk FS is more complicated.  For the $0\%$ and $0.27\%$ doped
samples, there is a bulk electron pocket inside the hexagram surface
state FS, with its size smaller for the $0.27\%$ doped sample. For
$0.67\%$ doped sample, the bulk electron pocket FS completely
vanishes, leaving no other FSs besides the surface state FS. For the
$0.9\%$ doped sample, on the contrary to $0\%$ and $0.27\%$ doped
samples, there are six leaves-like hole pockets outside the hexagon
surface stat FS, rising from the top of the bulk valence band(BVB).

\begin{figure}
\includegraphics[width=0.45\textwidth]{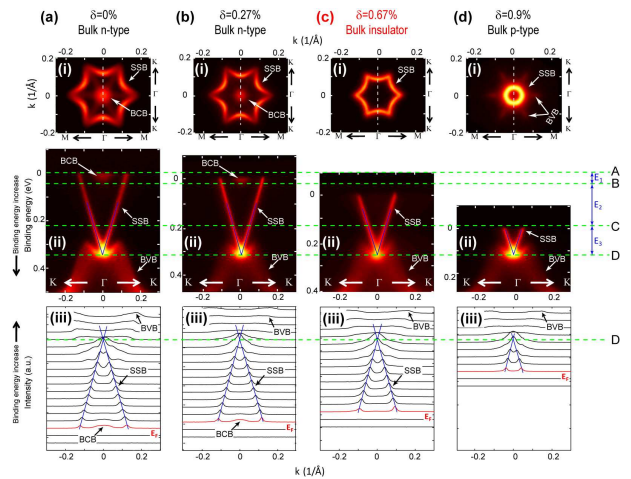}
\caption{\label{fig:epsart} (Color) Doping dependence of FSs and
$E_F$ positions.  (a)-(d): Measured FSs and band dispersions for
$0\%, 0.27\%, 0.67\%$ and $0.9\%$ nominally doped samples. Top row:
FS topology (symmetrized according to the crystal symmetry); Middle
row: Image plots of band dispersions along $K-\Gamma-K$ direction as
indicated by white dashed lines superimposed on the FSs in the top
row. Bottom row: Momentum distribution curve (MDC) plots of the raw
data. Definition of energy positions: A: $E_F$ position of undoped
$Bi_2Te_3$, B: BCB bottom, C: BVB top and D: Dirac point position,
determined by the intersection of the fitted linear dispersions.
Energy scale $E_1\sim E_3$ are defined in Fig. 1d.}
\end{figure}

The evolution of the band dispersion with doping is further
illustrated in the middle row of Fig. 2, where the Dirac points
(determined by the crossing point of the fit linear dispersions of
the ``V" shape surface states (SS)) from all four doping samples are
aligned to highlight the down-shift of $E_F$ due to the hole (Sn)
doping. In undoped $Bi_2Te_3$ (Fig. 2a(i), since $E_F$ lies at
0.34eV above the Dirac point and the BCB bottom is only 0.295eV
above it, $E_F$ intersects both the SS and BCB bands, resulting in a
hexagram and a snowflake FS pocket, respectively. In $0.27\%$ doped
sample (Fig. 2b(ii)), while $E_F$ is lowered by ~20meV due to Sn
doping, it still lies above the BCB minimum, thus both FS pockets
from SS and BCB still exist, although with smaller volumes as a
result of the lowered $E_F$. For the $0.67\%$ doped sample (Fig.
2c(ii)), significantly, $E_F$ is now further downshifted and resides
between the BCB bottom and the BVB top, therefore the FS pockets
associated with the bulk states completely vanish and the only FS
pocket left is the one originated from the SS(Fig. 2c(i)). This is
exactly what one should expect for a topological insulator.  With
further hole doing ($0.9\%$, Fig. 2d(ii)), $E_F$ is further shifted
downward and resides at 0.12eV above the Dirac point, slightly below
the BVB top which is 0.13eV above the Dirac point ($E_3$ in Fig.
1d(ii)). As a consequence, the bulk hole pockets emerge in the FS
map in Fig. 2d(i)) as the six ``leaves" outside the SS FS.  For all
doping levels, we have confirmed from photon energy dependence ARPES
that only the V-shape dispersion comes from the surface state while
other features are resulted from bulk states.

The third row of Fig. 2 shows the stack plots of raw momentum
distribution curves (MDCs) for the band dispersions shown in the
middle row, which fully support the observations and conclusions
above. The lineshape of the bulk states does not possess a sharp
peak because of the final state effect arisen from the kz
dispersion\cite{Miller}; on the contrary, the surface state
band(SSB) always exhibits sharp peak in the spectrum, again
confirming its 2-D character.

\begin{figure}
\includegraphics[width=0.45\textwidth]{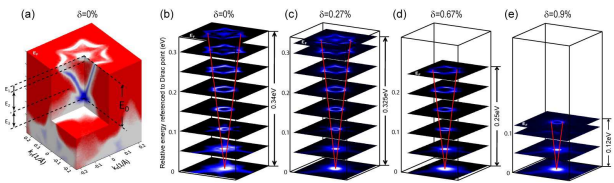}
\caption{\label{fig:epsart} (Color) (a) 3D illustration of the band
structures of undoped $Bi_2Te_3$, with the characteristic energy
scales $E_0\sim E_3$ defined in Fig. 1d. (b)$\sim$(d): Constant
energy contours of the band structure and the evolution of the
height of $E_F$ referenced to the Dirac point for the four dopings.
Red lines are guides to the eye that indicate the shape of the
constant energy band contours, and intersect at the Dirac point.}
\end{figure}

Unlike a simple circular Dirac cone, the observed surface state in
doped $Bi_2Te_3$ exhibits richer structure. As shown in Fig. 3, the
3D band structure (Fig. 3a) and the cross sections of the Dirac-like
dispersion at various binding energies are demonstrated (Fig.
3(b-e)). When approaching the Dirac point from $E_F$, the shape of
the SSB evolves gradually from a hexagram to a hexagon, then to a
circle of shrinking volume, and
 finally converges into a point, the
Dirac point, which is protected by the Kramers theorem. From the
doping evolution of the Fermi surface topology, band dispersion, and
the spectra lineshape shown above, we have found convincing evidence
that the $0.67\%$ Sn-doped $Bi_2Te_3$ is the long sought three
dimensional strong topological insulator with a single Dirac cone
and a large bulk band gap.

\begin{figure}
\includegraphics[width=0.45\textwidth]{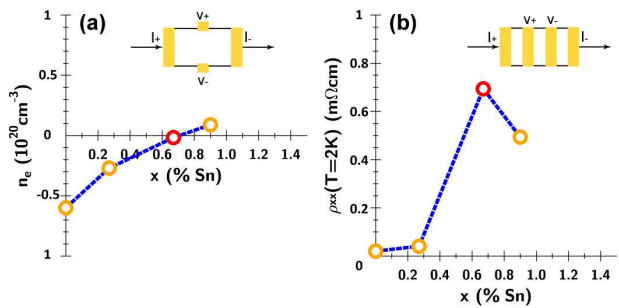}
\caption{\label{fig:epsart} (Color) Electric transport measurements
on samples of four different dopings. (a) Carrier density determined
by Hall coefficient measurement. Red symbol indicates the value for
$0.67\%$ doped sample, for which ARPES measurements show no bulk FS
pocket (see text). (b) Resistivity measured by four-probe methods.
Insets in (a), (b) show the measurement schematics.}
\end{figure}

The observations of ARPES are also supported by electric transport
measurements. From the Hall measurement (Fig.4a), an effective
three-dimensional carrier density (ne) is extracted as a function of
the Sn doping (Figure 4a). Evidently, at $\delta=0.67\%$ there is a
dramatic reduction in ne as the carrier type is inverted from n-type
to p-type. Similarly the in-plane resistivity(Fig. 4b) exhibits an
clear enhancement at $\delta=0.67\%$. The peak at this intermediate
doping indicates that the conductivity reduction is primarily due to
a dramatic decrease in the carrier density. The agreement between
the ARPES and transport measurements confirms that Sn doping drives
n-type $Bi_2Te_3$ into p-type, making $Bi_2Te_3$ an ideal parent for
the 3D topological insulator.

Our results on $Bi_2Te_3$ clearly show its distinction and
advantages over the previously studied material $Bi_{0.9}Sb_{0.1}$
and graphene: The single Dirac cone makes it the simplest model
system for studying the physics of topological insulators. In
particular, in order to observe the topological magneto-electric
effect, a thin magnetic layer needs to be coated on the surface to
break the time-reversal symmetry and create  a full insulating gap
at the Dirac point. Such effect is most likely to occur in a system
with a single Dirac cone on the surface\cite{Liu}. Furthermore, the
large bulk gap points to great potential for possible high
temperature spintronics applications on a 3D condensed matter system
which is easy to be realized with current standard semiconductor
technology.

\bigskip
\textbf{Acknowledgements} We thank W.S. Lee, K.J. Lai, B. Moritz,
C.X Liu for insightful discussions and C. Kucharczyk, L. Liu for
their assistance on crystal growth. This work is supported by the
Department of Energy, Office of Basic Energy Sciences under contract
DE-AC02-76SF00515; H.J.Z, Z.F and X.D acknowledge the supports by
the NSF of China, the National Basic Research Program of China, and
the International Science and Technology Cooperation Program of
China.

\end{document}